\documentclass{PoS}
\usepackage{subfigure}
\usepackage[fleqn,intlimits]{amsmath}

\title{SU(3)$_F$ in nonleptonic charm decays\thanks{DO-TH 13/30, QFET-2013-11, TTP13-038}}

\ShortTitle{SU(3)$_F$ in nonleptonic charm decays}

\author{Gudrun Hiller\\
	Institut f\"ur Physik, Technische Universit\"at Dortmund, D-44221 Dortmund, Germany\\
	E-mail: \email{gudrun.hiller@tu-dortmund.de}}

\author{Martin Jung\\
	Institut f\"ur Physik, Technische Universit\"at Dortmund, D-44221 Dortmund, Germany\\
	E-mail: \email{martin2.jung@tu-dortmund.de}}

\author{\speaker{Stefan Schacht}\\
	Institut f\"ur Physik, Technische Universit\"at Dortmund, D-44221 Dortmund, Germany\\
        Institut f\"ur Theoretische Teilchenphysik, Karlsruher Institut f\"ur Technologie, D-76128 Karlsruhe, Germany\\
        E-mail: \email{stefan.schacht@kit.edu}}

\abstract{
We present updated results of an SU(3)-flavor analysis of $D\rightarrow PP$  
decay data including linear breaking and no further assumptions. The global fit  is  
consistent with nominal $(\sim 30\%)$ SU(3)$_F$ breaking and
returns enhanced penguin (triplet) contributions. Their size is  
driven, in addition to $\Delta A_{CP}$, by  the CP asymmetries of
$D_0\rightarrow K_SK_S$, $D_s\rightarrow K_S\pi^+$ and $D_s\rightarrow K^+\pi^0$  decays. 
It is therefore especially important to improve these measurements.
}

\FullConference{The European Physical Society Conference on High Energy Physics\\
                18-24 July, 2013\\
                Stockholm, Sweden}

\begin{document}

\section{Introduction}

In 2011 and 2012, spectacular results indicated large CP violation in $D$ decays
\cite{Aaij:2011in,Collaboration:2012qw}.
Before the Moriond conference in March 2013, the world average for the difference of the
CP asymmetries $\Delta a_{CP}^{\mathrm{dir}}$ of $D^0\rightarrow K^+K^-$ and $D^0\rightarrow \pi^+\pi^-$ was $4.6\sigma$ away from zero.
This situation triggered a lot of theory papers 
\cite{Pirtskhalava:2011va, Feldmann:2012js, Brod:2012ud, Brod:2011re, Cheng:2012wr, Bhattacharya:2012ah, Franco:2012ck, Isidori:2011qw, Grossman:2012ry,Atwood:2012ac}, 
especially trying to explain the effect in new physics models,~\emph{e.g.}, 
\cite{Giudice:2012qq, Hiller:2012wf, Hochberg:2011ru, Altmannshofer:2012ur, Isidori:2012yx}. 
Recently, LHCb updated the previous measurement of $\Delta A_{CP}$ and performed a further measurement in an additional channel, resulting in 
\begin{align}
\Delta A_{CP} &= -0.0034 \pm 0.0018\,, & &\text{($D^*$ decay channel \cite{LHCb-CONF-2013-003})}  \\
\Delta A_{CP} &= 0.0049 \pm 0.0033\,,  & &\text{(semileptonic $B$ decay channel \cite{Aaij:2013bra})}  
\end{align}
where we added quadratically the statistical and systematic uncertainties. 
There is a tension of 2.2$\sigma$ \cite{Aaij:2013bra} between both results, which differ in sign.  
The current world average including these results is $\Delta a_{CP}^{\mathrm{dir}} = -0.00333 \pm 0.00120$ 
\cite{Aaij:2011in,Collaboration:2012qw,  LHCb-CONF-2013-003, Aaij:2013bra, CHARM13:Schwartz, Amhis:2012bh, Ko:2012px}, only
$2.8\sigma$ away from zero.\\
In the Standard Model (SM), one can write 
$\Delta a_{CP}^{\mathrm{dir}}\sim \left| P/T\right| \sin\delta\sin\gamma$ \cite{Brod:2011re,Grossman:2006jg} with the penguin 
over tree ratio $P/T$, the CKM angle $\gamma$ and a strong phase $\delta$. With respect to the tree amplitude, the penguin one
is CKM suppressed by $\lambda^4\sim 10^{-3}$ and a \lq\lq{}naive\rq\rq{} loop factor $\alpha_s/\pi\sim 0.1$. Altogether, this gives 
the rough estimate $\Delta a_{CP}^{\mathrm{dir}} \lesssim 10^{-4}$. 
The latter is however not reliable because $\alpha_s(m_c)$ is large and the expansion in $\Lambda_{\mathrm{QCD}}/m_c$ is 
not expected to converge fast. Consequently, it is unclear if a large $\Delta A_{CP}$ indicates new physics or is a QCD effect. In order to approach this problem we include additional observables which we relate by a symmetry principle.

\section{SU(3)$_F$ approach to nonleptonic $D$ decays} 

Besides $\Delta a_{CP}^{\mathrm{dir}}$ many other observables of $D\rightarrow PP$ decays have been measured. 
It is important to obtain a complete picture of all singly Cabibbo suppressed (SCS) CP asymmetries as well as 
SCS, Cabibbo-favored (CF) and doubly Cabibbo suppressed (DCS) branching ratios. 
In a data-driven way we use the approximate 
SU(3)$_F$ symmetry of QCD including linear breaking to relate 
several decay modes of $D\rightarrow PP$. For that, we analyze the operators of the effective Hamiltonian as well as initial and final states on their 
SU(3)$_F$ representations. Subsequently, we use the Wigner-Eckart theorem in order to obtain the corresponding reduced matrix elements, all of which we fit from data only. 
The amplitudes of SCS, CF and DCS amplitudes can then be written as
\begin{align}
\mathcal{A}(d) &= \Sigma \left(
	\sum_{i,k} c_{d;ik} A_i^k + 
	\sum_{i,j} c_{d;ij} B_i^j
	\right),  & &\mathrm{(SCS)}  \\
\mathcal{A}(d) &= V_{cs}^* V_{ud} \left(
	\sum_{i,k} c_{d;ik} A_i^k +
	\sum_{i,j} c_{d;ij} B_i^j
	\right), & &\mathrm{(CF)}  \\
\mathcal{A}(d) &= V_{cd}^* V_{us} \left(
	\sum_{i,k} c_{d;ik} A_i^k + \sum_{i,j} c_{d;ij} B_i^j
	\right), & &\mathrm{(DCS)}
\end{align}
with $\Sigma\equiv\left( V_{cs}^* V_{us} - V_{cd}^* V_{ud}\right)/2$, the SU(3)$_F$ limit matrix elements 
$A_i^k$, the SU(3)$_F$-breaking matrix elements $B_i^j$ and Clebsch-Gordan coefficients $c_{d;ik}$. CP violation is induced by 
the interference of terms coming with the CKM factors $\Sigma$ and $\Delta\equiv \left(V_{cs}^*V_{us} + V_{cd}^* V_{ud}\right)/2$.
The dominant contribution to CP-violating observables is coming from penguin matrix elements $A_1^3$ and $A_8^3$ that come with 
coefficients $c_{d;(1,8)3} \propto \Delta$.
The SU(3)$_F$ decomposition and the corresponding Clebsch-Gordan coefficients can be found in \cite{Hiller:2012xm}.\\
Here we present, as part of recent work \cite{Hiller:2013prep}, an update of our SU(3)$_F$ study \cite{Hiller:2012xm} of 
the presently available data on two-body charm decays to kaons and pions. 
Since fall of 2012, several new measurements appeared, which modify the world averages of 
the CP asymmetries $\Delta a_{CP}^{\mathrm{dir}}$, 
$\Sigma a_{CP}^{\mathrm{dir}}\equiv a_{CP}^{\mathrm{dir}}(D^0\rightarrow K^+K^-)+a_{CP}^{\mathrm{dir}}(D^0\rightarrow\pi^+\pi^-)$, 
$a_{CP}^{\mathrm{dir}}(D_s\rightarrow K_S\pi^+)$, the strong phase $\delta_{K\pi}$ as well as the branching ratio of the 
CF decay $D_s\rightarrow K_SK^+$. For a comparison with the values used in \cite{Hiller:2012xm} as well as detailed references see Table~\ref{tab:data}. 
Besides the decrease of $\Delta a_{CP}^{\mathrm{dir}}$ and $a_{CP}^{\mathrm{dir}}(D_s\rightarrow K_S\pi^+)$, the measurement 
of $\Sigma a_{CP}^{\mathrm{dir}}$ gained precision and its central value approached zero. The latter is in accord with the $U$-spin limit 
relation $\Sigma a_{CP}^{\mathrm{dir}} = 0$.

The fits shown are performed using the \texttt{NLopt} code \cite{NLopt} with the algorithms that can be 
found in \cite{NLopt, conn1991globally, birgin2008improving, Rowan1990}.

\section{Extraction of SU(3)$_F$ breaking and penguins from present data}
\label{sec:su3flavor}

\begin{figure}[!tp]
\centering{
\subfigure[SU(3)$_F$ breaking.\label{fig:deltaX-deltaXprime}]{
	\includegraphics[width=0.42\textwidth]{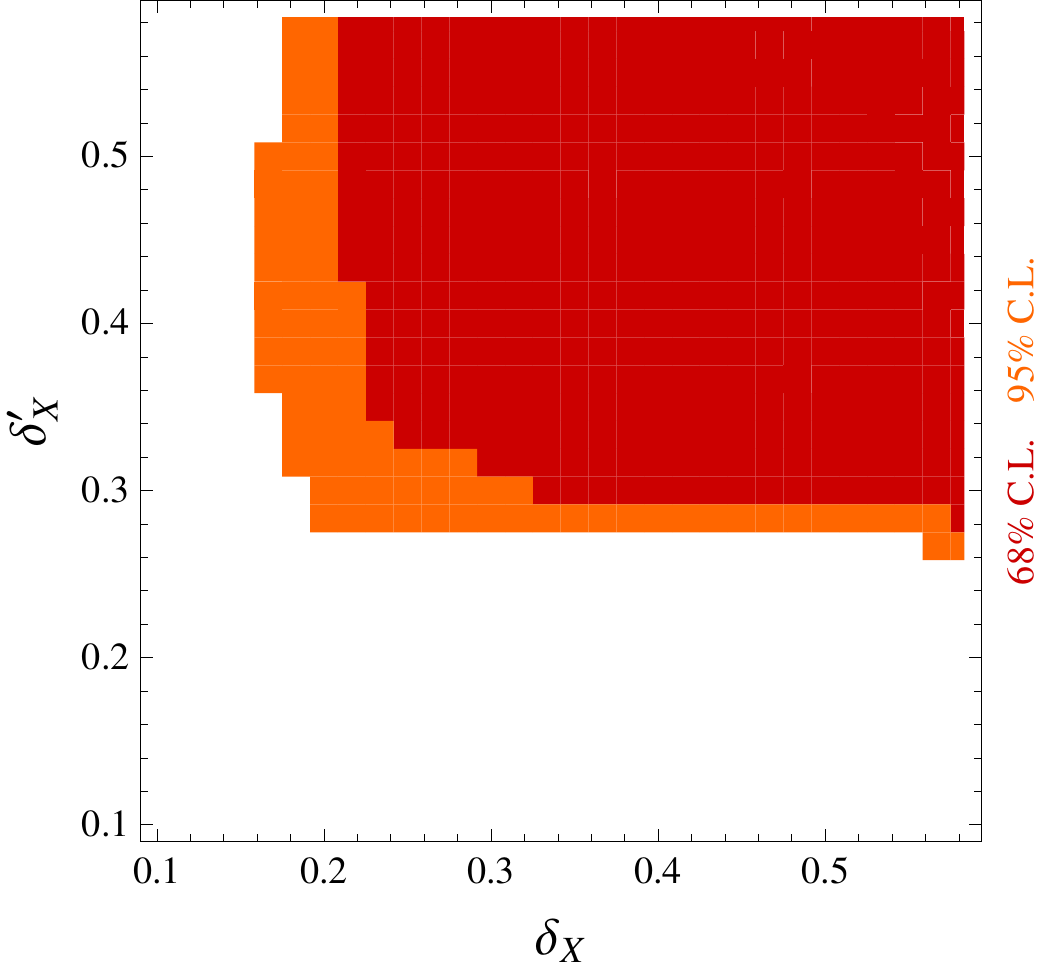}
}
\subfigure[Relative size of Penguins.\label{fig:delta3prime-delta3}]{
	\includegraphics[width=0.42\textwidth]{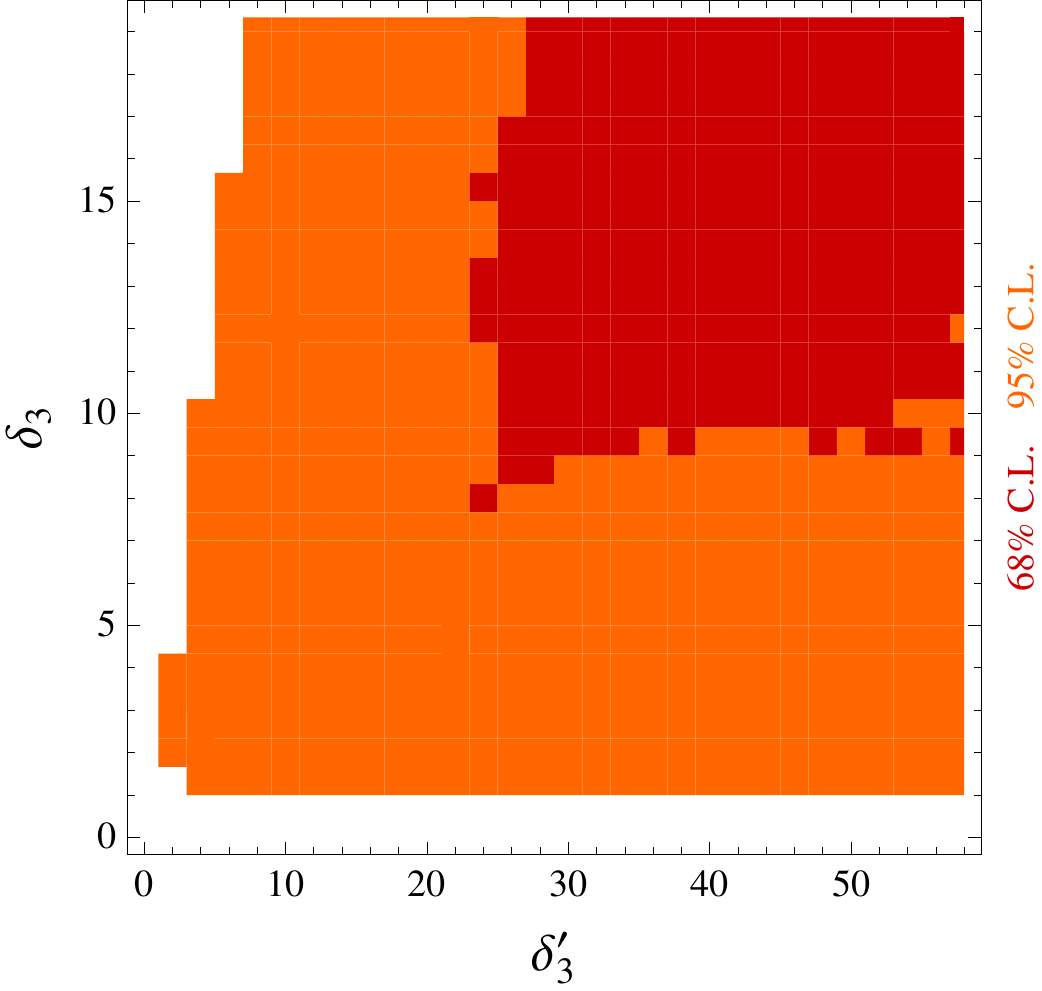}
}
\caption{Fit results for SU(3)$_F$ breaking and penguin enhancement, taking all present data given in Table~II of \cite{Hiller:2012xm} and updated in Table~\protect\ref{tab:data} into account. The red (orange) regions are allowed at 68\% and 95\% C.L., respectively, see text for details. Figures taken from \cite{Hiller:2013prep}.}
}
\end{figure}

In order to quantify the SU(3)$_F$ breaking and the size of the penguins we introduce the following 
complementary measures \cite{Hiller:2012xm}:
\begin{align}
\delta_X &= \frac{
\mathrm{max}_{ij} \vert B_i^j\vert
}{
\mathrm{max}\left(\vert A_{27}^{15}\vert, \vert A_8^{\bar{6}}\vert, \vert A_8^{15}\vert\right)
}\,, & 
\delta'_X &= \mathrm{max}_d \left|\frac{\mathcal{A}_X(d)}{\mathcal{A}(d)}\right|\,, \label{eq:measures-1}  \\
\delta_3 &= \frac{
	\mathrm{max}\left(\vert A_1^3\vert, \vert A_8^3\vert\right)
}{
	\mathrm{max}\left(\vert A_{27}^{15}\vert, \vert A_8^{\bar{6}}\vert, \vert A_8^{15}\vert \right) 
}\,, &
\delta'_3 &= \mathrm{max}_d \left|
	\frac{
	c_{d;1\, 3} A_1^3 + c_{d;8\, 3} A_8^3
}{
	c_{d;27\, 15} A_{27}^{15} + c_{d;8\, \bar{6}} A_8^{\bar{6}} + c_{d;8\, 15} A_8^{15} 
}\right|\,. \label{eq:measures-2} 
\end{align}
Here, $\mathcal{A}_X(d)$ is the SU(3)$_F$-breaking part of the amplitudes, \emph{i.e.},~$\mathcal{A}(d)$ with $A_i^k=0$.
In the definition of the measures $\delta'_X$ and $\delta'_3$ we do not take into account the decay $D^0\rightarrow K_SK_S$. 
The latter would introduce a bias, since, in contrast to all other considered decays, its SU(3)$_F$ limit is CKM-suppressed $\propto\Delta$. 
The ratios of SU(3)$_F$ matrix elements $\delta_X$ and $\delta_3$ do not take 
into account effects of small Clebsch-Gordan coefficients in front of the respective matrix elements. On the 
other hand, the ratios of parts of amplitudes, $\delta^\prime_X$ and $\delta^\prime_3$ do not take into account possible large cancellations between 
different summands. It is therefore essential to study both measures to achieve the full information on the 
SU(3)$_F$ breaking and the penguin enhancement.

The truncation of the perturbative SU(3)$_F$ expansion at the next to leading order is only sensible if the SU(3)$_F$ breaking is not too large. Therefore at first we validate our ansatz.
In Fig.~\ref{fig:deltaX-deltaXprime}, we show our fit results for the SU(3)$_F$ breaking with present data as summarized in Table~II of \cite{Hiller:2012xm} and 
updated in Table~\ref{tab:data}. 
Already an SU(3)$_F$ breaking of $\sim 30\%$ can describe the data reasonably well, while larger SU(3)$_F$ breaking cannot be excluded from data only. 
This is true independent of the measure $\delta_X$ or $\delta^\prime_X$. 
Therefore, the SU(3)$_F$ ansatz is justified for nonleptonic charm decays. 
Consequently, we can proceed with our analysis and study the size of the penguins. 

\begin{figure}[!tp]
\centering{
\subfigure[All data.\label{fig:delta3prime-delta3-zoom}]{
	\includegraphics[width=0.42\textwidth]{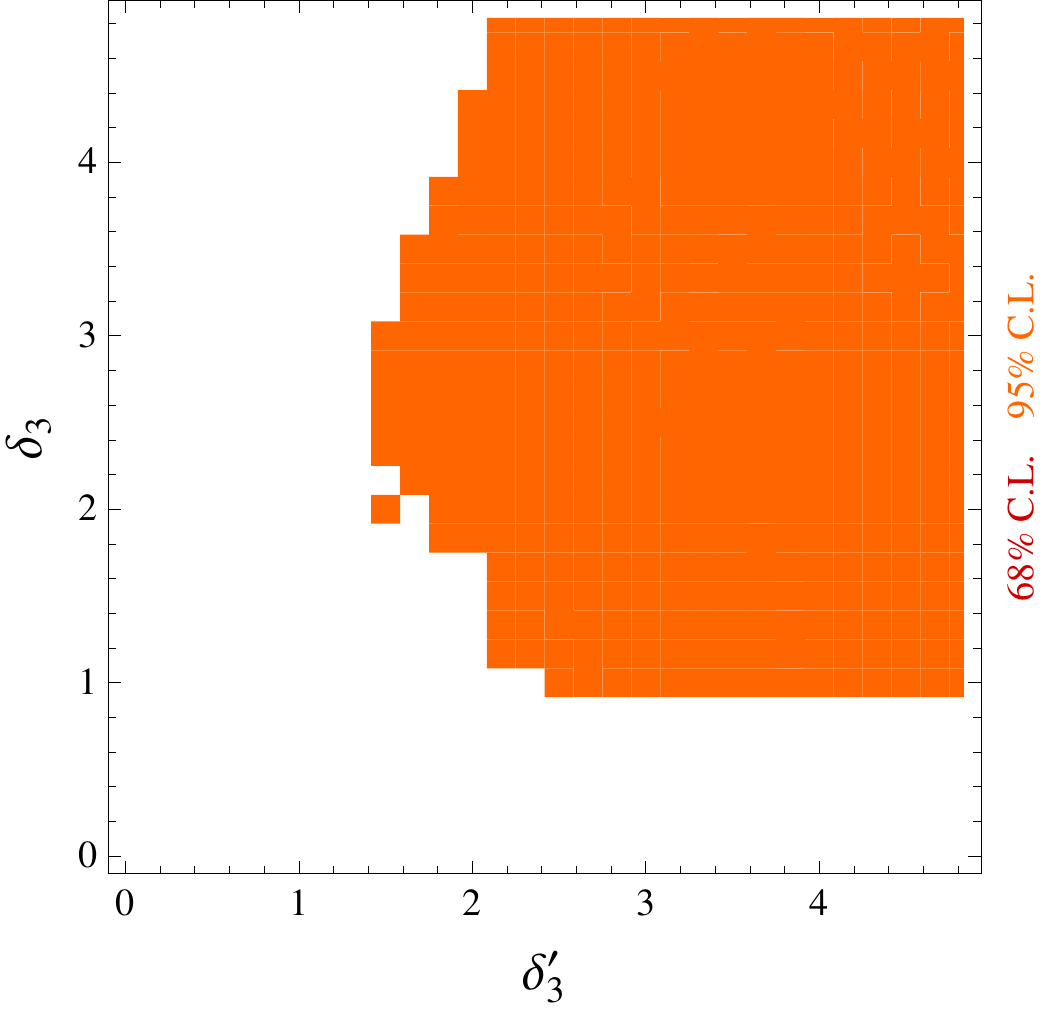}
	}
\subfigure[Without large CP asymmetries.\label{delta3prime-delta3-wo-large-cp-asym}]{
	\includegraphics[width=0.42\textwidth]{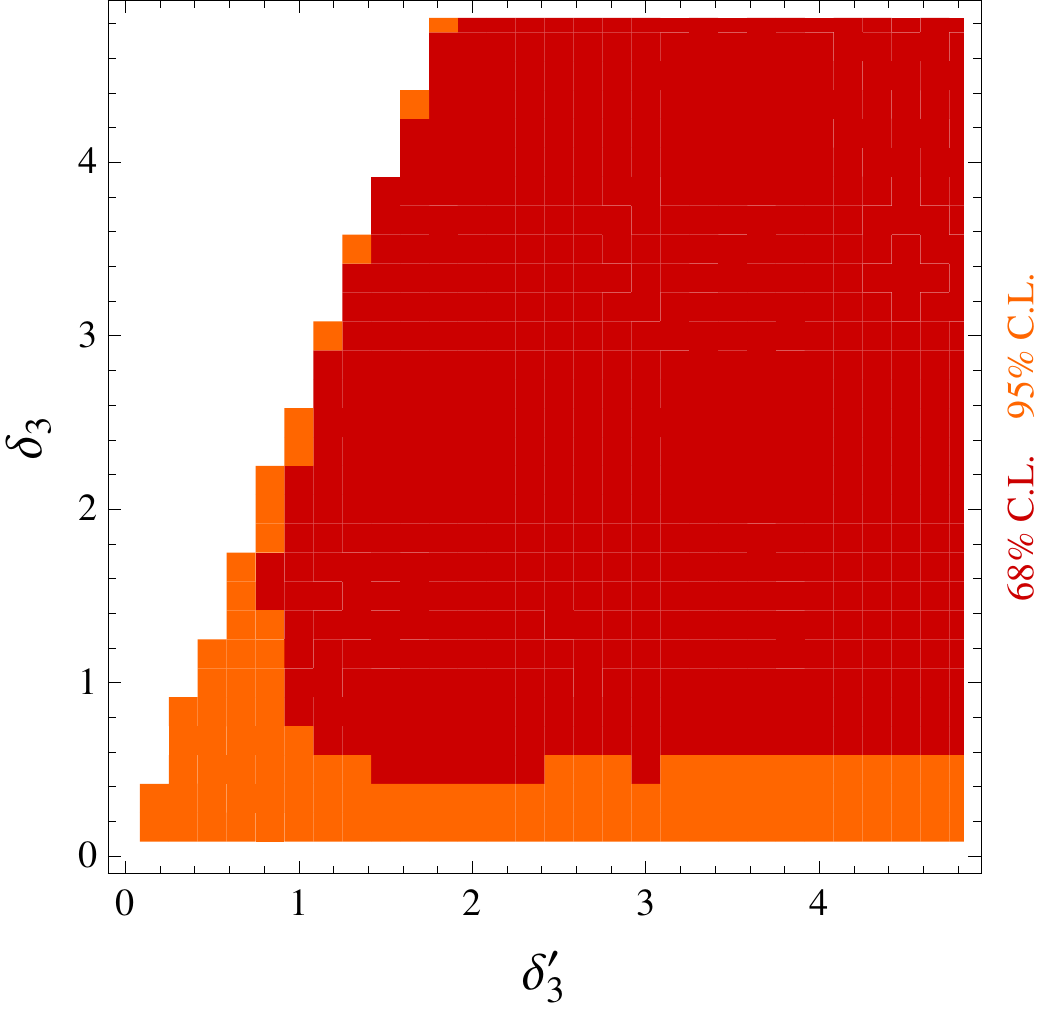}
	}
\caption{Zoom of fit results in the $\delta'_3$--$\delta_3$ plane. Left: taking all data into account. Right: Without the CP asymmetries of the decays $D^0\rightarrow K_SK_S$, $D_s\rightarrow K_S\pi^+$ and $D_s\rightarrow K^+\pi^0$. 
The red (orange) regions are allowed at 68\% and 95\% C.L., respectively, see text for details.
Figures taken from \cite{Hiller:2013prep}.}}
\end{figure}

Our fit results to the current data for the relative size of the penguins, or more specifically, the triplet matrix elements which come with $\Delta$,
are shown in Fig.~\ref{fig:delta3prime-delta3}. The generic SM expectation is $\delta_3^{(\prime)} \sim P/T \sim 0.1$, while
values of $\delta_3^{(\prime)} \sim 1$ are generally regarded as an enhancement. 
As can be seen from Fig.~\ref{fig:delta3prime-delta3}, the allowed 68\%~C.L. region is located at very high values 
of $\delta_3^{(\prime)}$. The border of the 95\%~C.L. region can be identified in the zoom of the $\delta'_3$--$\delta_3$ plane 
which is shown in Fig.~\ref{fig:delta3prime-delta3-zoom} and which shows the fit result $\delta^{(\prime)}_3\gtrsim \mathcal{O}(1)$.
The reason is the following: besides $\Delta a_{CP}^{\mathrm{dir}}$, there are additional CP asymmetries in the global
analysis that are measured with largish central values and generate the need for a penguin enhancement. This is illustrated in Fig.~\ref{delta3prime-delta3-wo-large-cp-asym}, where 
we take for comparison the observables $a_{CP}^{\mathrm{dir}}(D^0\rightarrow K_SK_S)$, 
$a_{CP}^{\mathrm{dir}}(D_s\rightarrow K_S\pi^+)$ and 
$a_{CP}^{\mathrm{dir}}(D_s\rightarrow K^+\pi^0)$ out of the fit. Without them, also $\delta^{(\prime)}_3\lesssim \mathcal{O}(1)$ becomes allowed in the fit. 
The tendency towards large values for the penguins in the global analysis highlights
the importance of future improved measurements of these asymmetries. These will enable us to better constrain the size of the penguins. 

\begin{table}[t]
\begin{center}
\begin{tabular}{cccr}
\hline\hline
Observable            &  Average Nov 2013  &  Average Nov 2012 \cite{Hiller:2012xm}  &  References  \\\hline\hline
$\Delta a_{CP}^{\mathrm{dir}}(K^+K^-,\pi^+\pi^-)$   & $-0.00333 \pm 0.00120$ &  $-0.00678\pm 0.00147$  & \cite{CHARM13:Schwartz},\cite{Aaij:2011in, Collaboration:2012qw, LHCb-CONF-2013-003, Aaij:2013bra} \\ & & & \cite{Amhis:2012bh, Ko:2012px, Aubert:2007if, Staric:2008rx}  \\
$\Sigma a_{CP}^{\mathrm{dir}}(K^+K^-, \pi^+\pi^-)$     &       $+0.00008 \pm 0.00228$ & $+0.0014\pm 0.0039$  & 
$^\dagger$\cite{Aaij:2011in, Collaboration:2012qw, Ko:2012px,Aubert:2007if,Aaltonen:2011se}   \\
$a_{CP}^{\mathrm{dir}}(D_s\rightarrow K_S \pi^+)$  & $+0.011 \pm 0.007$  & $+0.028\pm 0.015$  & $^\dagger$\cite{Mendez:2009aa, Lees:2012jv,Ko:2010ng,Aaij:2013ula}   \\
$a_{CP}^{\mathrm{ind}}$   &   $+0.00015 \pm 0.00052$  &  $-0.00027\pm 0.00163$  &    \cite{CHARM13:Schwartz}   \\
$\delta_{K\pi}$   &  $(11.7 \pm 10.2)^{\circ}$  & $(21.4 \pm 10.4)^{\circ}$  &    $^\ddagger$\cite{CHARM13:Schwartz}    \\
$\mathcal{B}(D_s\rightarrow K_S K^+  )   $   & $   \left(1.50\pm  0.05\right)\cdot10^{-2}  $ & $(1.45\pm 0.05)\cdot 10^{-2}  $ & $^\dagger$\cite{Zupanc:2013byn,Onyisi:2013bjt}  \\\hline\hline
\end{tabular}
\end{center}
\caption{Updates of averages of experimental measurements compared to the status in Nov 2012 given in \cite{Hiller:2012xm}. $^\dagger$Our average where we added systematic and statistical error quadratically. 
$^\ddagger$Uncertainties calculated by symmetrization. Table adapted from \cite{Hiller:2012xm} and \cite{Hiller:2013prep}. \label{tab:data}}
\end{table}

\section{Future data}

There are plans to measure the CP asymmetry $a_{CP}^{\mathrm{dir}}(D^0\rightarrow \pi^0 \pi^0)$ in the future with 
a $\sim10$ times smaller uncertainty than is the case at present \cite{CHARM13:Schwartz}.  
We study therefore a hypothetical future data scenario, assuming that $a_{CP}^{\mathrm{dir}}(D^0\rightarrow \pi^0 \pi^0)$ is measured as 
\begin{align}
a_{CP, \mathrm{future}}^{\mathrm{dir}}(D^0\rightarrow \pi^0\pi^0) = 0.000\pm 0.006\,, \qquad \text{(future data)} 
\end{align}
where the uncertainty is motivated by the prospect given by Belle \cite{CHARM13:Schwartz}.
The corresponding 95\%~C.L. contour of the penguin enhancement differs hardly from the one shown in 
Fig.~\ref{fig:delta3prime-delta3-zoom}. This result can be understood from counting the degrees of freedom in the fit: in order to determine the two complex 
penguin matrix elements that induce CP violation, at least 4 significant measurements of SCS CP asymmetries are needed. 
This means that, while present uncertainties leave much room for possible enhancements, even in the 
pessimistic scenario of no CP violation showing up in $D^0\rightarrow \pi^0\pi^0$ decays, the search for CP violation in SCS decays remains interesting.

\section{Conclusion}
\label{conclusion} 

The global SU(3)$_F$ fit of two-body charm decays to kaons and pions shows that an SU(3)$_F$ breaking of $\sim 30\%$ 
suffices to describe the data. Furthermore, it reveals a tendency towards large values for the penguins. 
The penguin enhancement has decreased somewhat with respect to an earlier study \cite{Hiller:2012xm}, especially due to 
new experimental results on $\Delta a_{CP}^{\mathrm{dir}}$. Nevertheless, taking all observables into account, the overall 
characteristics found in \cite{Hiller:2012xm} persist.   
The observed penguin enhancement is driven, in addition to $\Delta a_{CP}^{\mathrm{dir}}$, by the CP asymmetries 
$a^{\mathrm{dir}}_{CP}(D^0\rightarrow~K_SK_S)$, 
$a^{\mathrm{dir}}_{CP}(D_s\rightarrow~K_S\pi^+)$ and 
$a^{\mathrm{dir}}_{CP}(D_s\rightarrow~K^+\pi^0)$. These are not measured significantly on their own, so that
at present the situation is inconclusive regarding the  
validity of the Standard Model. It is therefore very important to improve these measurements. 
Furthermore, the tendency to large triplet matrix elements remains assuming a hypothetical but realistic \cite{CHARM13:Schwartz} 
measurement of the CP asymmetry in $D^0\rightarrow\pi^0 \pi^0$ at the few-permille level but consistent with zero.

We summarize here the following characteristics of $D\rightarrow PP$ decays, that can serve as a guide 
for future measurements:
\begin{itemize}
\item The CKM-leading contribution to the decay $D^0\rightarrow K_SK_S$ comes first into play when taking SU(3)$_F$ breaking into account. Therefore its CP asymmetry is enhanced  with respect to $\Delta a_{CP}^{\mathrm{dir}}$ as $\mathcal{O}(1/\delta_X)$ in the SM as well as in generic new physics models \cite{Hiller:2012xm,Atwood:2012ac}.
\item Due to isospin symmetry in the SM to very good precision $a_{CP}^{\mathrm{dir}}(D^+\rightarrow\pi^0\pi^+) = 0$. The violation of this rule would be 
	a smoking gun for $\Delta I=3/2$ new physics \cite{Hiller:2012xm,Grossman:2012eb}.
\item The $\mathcal{O}(1)$ breaking of the U-spin relations
	\begin{align}
	a^{\mathrm{dir}}_{CP}(D^0\rightarrow K^+ K^-) + a^{\mathrm{dir}}_{CP}(D^0\rightarrow \pi^+\pi^-)     &= 0\,,\\
	a^{\mathrm{dir}}_{CP}(D^+\rightarrow \bar{K}^0 K^+) + a^{\mathrm{dir}}_{CP}(D_s\rightarrow K^+\pi^0) &= 0\,, \label{eq:sum-rule-02}
	\end{align}
	beyond the amount of SU(3)$_F$ breaking would be a sign of new physics \cite{Hiller:2012xm}. 
	Especially with regard to Eq.~(\ref{eq:sum-rule-02})
	there is much space left for an improvement in experimental precision. 
\item The present tendency towards a penguin enhancement is driven by the CP asymmetries $\Delta a_{CP}^{\mathrm{dir}}$, $a_{CP}^{\mathrm{dir}}(D^0\rightarrow K_SK_S)$, $a_{CP}^{\mathrm{dir}}(D_s\rightarrow K_S\pi^+)$ and $a_{CP}^{\mathrm{dir}}(D_s\rightarrow K^+\pi^0)$ \cite{Hiller:2012xm}.
\end{itemize} 
SU(3)$_F$ analysis allows improved $D\rightarrow PP$ measurements to provide more precise information on the borders of the SM and new physics in $\vert \Delta C\vert=1$ processes.

\section*{Acknowledgments}

StS gratefully acknowledges support by the DFG Research Unit FOR 1873 
\lq\lq{}Quark Flavour Physics and Effective Field Theories\rq\rq{}.

\end{document}